\RequirePackage{fix-cm}
\documentclass[twocolumn]{svjour3}          
\smartqed  
\usepackage{graphicx, color}
\usepackage{dcolumn}
\usepackage{bm}
\usepackage{ulem}
\usepackage{xspace}
\usepackage[utf8]{inputenc}
\usepackage[T1]{fontenc}
\usepackage{nicefrac}

\newcolumntype{d}{D{.}{.}{-1}}    
\newcolumntype{b}{D{(}{\ (}{-1}}  

\newcommand{\Ca}[1]{\ensuremath{^{#1}}Ca\ensuremath{^{+}}}

\newcommand{\kms}[1]{\ensuremath{K_{#1}}}
\newcommand{\ksms}[1]{\ensuremath{K_{{#1}\mathrm{SMS}}}}
\newcommand{\knms}[1]{\ensuremath{K_{{#1}\mathrm{NMS}}}}
\newcommand{\fs}[1]{\ensuremath{F_{#1}}}
\newcommand{\drsq}{\ensuremath{\delta\!\left \langle r_c^2 \right \rangle}}

\newcommand{\GHzamu}{GHz$\cdot$amu\xspace}
\newcommand{\MHzfmsq}{\textrm{MHz$\cdot$fm$^{-2}$}}
\newcommand{\fmsq}{\ensuremath{\textrm{fm}^{2}}}

\def\fm#1{\ifmmode #1 \else $#1$\fi}
\def\ket#1{{%
  \ifmmode |\,#1\,\rangle \else $|\,#1\,\rangle$\fi}}
\def\bra#1{{%
  \ifmmode \langle\,#1\,| \else $\langle\,#1\,|$\fi}}
\def\braket#1#2{{%
  \ifmmode \langle\,#1\,|\,#2\,\rangle \else $\langle\,#1\,|\,#2\,\rangle$\fi}}
\def\expect#1{{%
  \ifmmode \langle\,#1\,\rangle \else $\langle\,#1\,\rangle$\fi}}

\def\dsoh{\fm{{}^2\mathrm{S}_{\nicefrac{1}{2}}}\xspace}
\def\dpoh{\fm{{}^2\mathrm{P}_{\nicefrac{1}{2}}}\xspace}
\def\dpth{\fm{{}^2\mathrm{P}_{\nicefrac{3}{2}}}\xspace}
\def\ddth{\fm{{}^2\mathrm{D}_{\nicefrac{3}{2}}}\xspace}
\def\ddfh{\fm{{}^2\mathrm{D}_{\nicefrac{5}{2}}}\xspace}

\def\dsohdpoh{\dsoh\fm{\rightarrow}\enskip\dpoh\xspace}
\def\dsohdpth{\dsoh\fm{\rightarrow}\enskip\dpth\xspace}
\def\ddthdpoh{\ddth\fm{\rightarrow}\enskip\dpoh\xspace}
\def\ddfhdpth{\ddfh\fm{\rightarrow}\enskip\dpth\xspace}

\journalname{Applied Physics B}

\begin{document}

\title{Unexpectedly large difference of the electron density at the nucleus in the $ 4p\, ^2\mathrm{P}_{\nicefrac{1}{2},\nicefrac{3}{2}}$ fine-structure doublet of Ca$^+$}

\titlerunning{Unexpectedly large difference of $\left|\psi(0)\right|^2$ in the $ 4p\, ^2\mathrm{P}_{\nicefrac{1}{2},\nicefrac{3}{2}}$ fine-structure doublet of Ca$^+$}        

\author{C.~Shi$^1$               \and
        F.~Gebert$^1$                  \and
        C.~Gorges$^2$               \and
        S.~Kaufmann$^2$             \and
        W.~N\"ortersh\"auser$^{2}$  \and
        B.~K.~Sahoo$^{3}$            \and
        A.~Surzhykov$^{1,4}$          \and
        V.A.~Yerokhin$^{1,5}$           \and
        J.C.~Berengut$^{6}$ \and
        F.~Wolf$^{1}$ \and
        J.~C.~Heip$^{1}$ \and
        P.~O.~Schmidt$^{1,7}$
       }

\institute{\\
        $^{1}$Physikalisch-Technische Bundesanstalt, 38116 Braunschweig, Germany\\
        $^{2}$Institut f\"ur Kernphysik,     Technische Universit\"at Darmstadt, 64289 Darmstadt, Germany \\
        $^{3}$Atomic, Molecular and Optical Physics Division, Physical Research Laboratory, Navrangpura, Ahmedabad 380009, India\\
        $^{4}$Technische Universit\"at Braunschweig, 38106 Braunschweig, Germany\\
        $^{5}$Center for Advanced Studies, Peter the Great St. Petersburg Polytechnic University, 195251 St. Petersburg, Russia\\
        $^{6}$School of Physics, University of New South Wales, Sydney, New South Wales 2052, Australia\\
        $^{7}$Institut f\"ur Quantenoptik, Leibniz Universit\"at Hannover, 30169 Hannover, Germany
}

\maketitle

\authorrunning{C.~Shi et al.} 

\date{Received: date / Accepted: date}

\begin{abstract}
We measured the isotope shift in the \dsohdpth (D2) transition in singly-ionized calcium ions using photon recoil spectroscopy. The high accuracy of the technique enables us to resolve the difference between the isotope shifts of this transition to the previously measured isotopic shifts of the \dsohdpoh (D1) line. This so-called splitting isotope shift  is extracted and exhibits a clear signature of field shift contributions. From the data we were able to extract the small difference of the field shift coefficient and mass shifts between the two transitions with high accuracy. This $J$-dependence is of relativistic origin and can be used to benchmark atomic structure calculations. As a first step, we use several \textit{ab initio} atomic structure calculation methods to provide more accurate values for the field shift constants and their ratio. Remarkably, the high-accuracy value for the ratio of the field shift constants extracted from the experimental data is larger than all available theoretical predictions.
\end{abstract}


\section{\label{sec:introduction} Introduction}

The study of isotopic shifts in atomic systems has a long history \cite{Breit1958} and a profound understanding of the isotope shift and theoretical calculations of the atomic properties is important in many applications. These reach from the extraction of nuclear properties from atomic spectra to applications in astronomy and fundamental physics. The spectrum of atoms and ions encodes information that provides a key to the ground-state properties of nuclei \cite{Campbell2016,Blaum2013}, small parity-violating effects caused by the weak interaction \cite{Dzuba2005}, for unitarity tests of the Cabibbo-Kobayashi-Maskawa (CKM) matrix \cite{Mane2011}, or for probing the Higgs coupling between electrons and quarks \cite{Delaunay2016}.\\
Light appearing on earth from stars at large distances is red shifted and can provide information about the spectra of atoms in ancient times and whether there have been changes, for example from a variation of the fine structure constant $\alpha$. However, isotopic composition can also contribute to the observed shifts since the isotopes have different resonance frequencies, but this so-called isotope shift is usually not resolved \cite{Kozlov2004,berengut_atomic_2011,murphy_laboratory_2014}.
In this respect, isotope shift calculations became recently an important topic with the goal to either determine the influence of the isotopic abundance of the observed species on the analysis for a change in $\alpha$ \cite{Kozlov2004,Korol2007}, or to provide information of the isotopic composition in the ancient times of the universe.\\
Calcium is an element of considerable interest for many of the cases mentioned above. For example the isotope shift information in the $3d \; ^2D_J \rightarrow 4p \; ^2P_J$ infrared triplet \cite{Noertershaeuser1998} led to the discovery of an anomalous isotopic composition in mercury-manganese (HgMn) stars, in which the isotopic Ca ratio in the stellar atmosphere is dominated by $^{48}$Ca \cite{Castelli2004,Cowley2005}. Isotope shifts in other calcium transitions have been studied, e.g., to extract nuclear charge radii along the long chain of isotopes \cite{Palmer1984,Martensson1992,Vermeeren1992,GarciaRuiz2016} or to perform ultra-trace analysis using isotope selective resonance ionization \cite{LuWendt2003}. Moreover, the calcium ion is a workhorse in the field of quantum-optical applications and transition frequencies and isotope shifts of stable isotopes were measured with high accuracy \cite{Wolf2008,Wolf2009,Wan2014,Gebert2015} and supported on-line studies of exotic isotopes since they serve as calibration points \cite{Gorges2015,GarciaRuiz2016}.

The calcium isotopic chain is quite unique since it contains two stable doubly magic isotopes \Ca{40,48} that have practically identical mean-square charge radii even though they are 20\,\% different in mass. This has been established using a variety of techniques, i.e. elastic electron scattering \cite{Emrich1983}, muonic atom spectroscopy \cite{Wohlfahrt1981} as well as optical isotope shift data, e.g. \cite{Martensson1992} (for a synopsis see e.g. \cite{Rebel1979} and references therein). The negligible change in nuclear charge radius between the two isotopes allows for a cleaner separation of mass and field shifts than in most other multi-electron systems. The high-precision data presented here will provide important benchmarks for improved calculations of this reference system.

The extraction of nuclear parameters from atomic spectra is strongly facilitated by  atomic structure calculations.
Despite being a lighter system with only 19 electrons, isotope shifts in the singly ionized calcium (Ca$^+$) have not been studied rigorously. High precision calculations of magnetic dipole and electric quadrupole hyperfine structure
constants have been performed in this ion using an all order relativistic many-body theory in the coupled-cluster (RCC) theory framework \cite{Sahoo2009}.
However, calculations of field shift and mass shift constants that are required to estimate isotope shifts have not been performed at
the same level of accuracy yet.

Here we present a high precision absolute frequency measurement of the D2 line of \Ca{40} with 100~kHz accuracy, representing a
five-fold improvement over previous results \cite{Wolf2009}. Isotope shift measurements of this transition with the same resolution
are compared with a measurement of the D1 line \cite{Gebert2015}. A clear signature of
field shift contributions to the splitting isotope shift are observed. 
To explain this finding, we have also performed several {\it ab initio} calculations of field shift constants employing a hydrogenic method, a mean-field method using Dirac-Fock (DF) equation, and state-of-the-art atomic structure calculations.

\section{\label{sec:setup} Experimental setup}

The isotope shift in the \dsohdpth (D2) transition of even calcium isotopes was measured by photon recoil spectroscopy, as described in detail in references \cite{Wan2014,Gebert2015}. In brief, we trap a singly charged $^{25}$Mg$^{+}$ ion together with the calcium isotope under investigation in a linear Paul trap. The $^{25}$Mg$^{+}$ ion is used to sympathetically cool the axial normal mode of the two-ion-crystal to the ground state \cite{Wan2015}. To probe the transition, a series of 70 pulses of the spectroscopy laser with a pulse length of 125\,ns, synchronized to one of the motional frequencies of the two-ion crystal are applied. Each spectroscopy laser pulse is followed by 200~ns short repump pulses on the \ddthdpoh and \ddfhdpth
transitions at 866\,nm and 854\,nm, respectively. Recoil kick upon photon absorption on the spectroscopy transition results in excitation of nearly coherent motion. This motional excitation is mapped into an electronic excitation using a stimulated rapid adiabatic passage (STIRAP) pulse on the $^{25}$Mg$^{+}$ ion \cite{Gebert2016}. The high photon sensitivity of this technique provides a large signal-to-noise ratio, resulting in a resolution of about 100\,kHz in less than 15\,min of averaging time. As a consequence of the smallness of systematic effects, the accuracy of absolute frequency measurements is also about 100\,kHz. Isotope shift measurements benefit from further suppression of systematic effects, since most of them are common to all isotopes.
\begin{figure}[h]
\centering
\includegraphics[width=\linewidth]{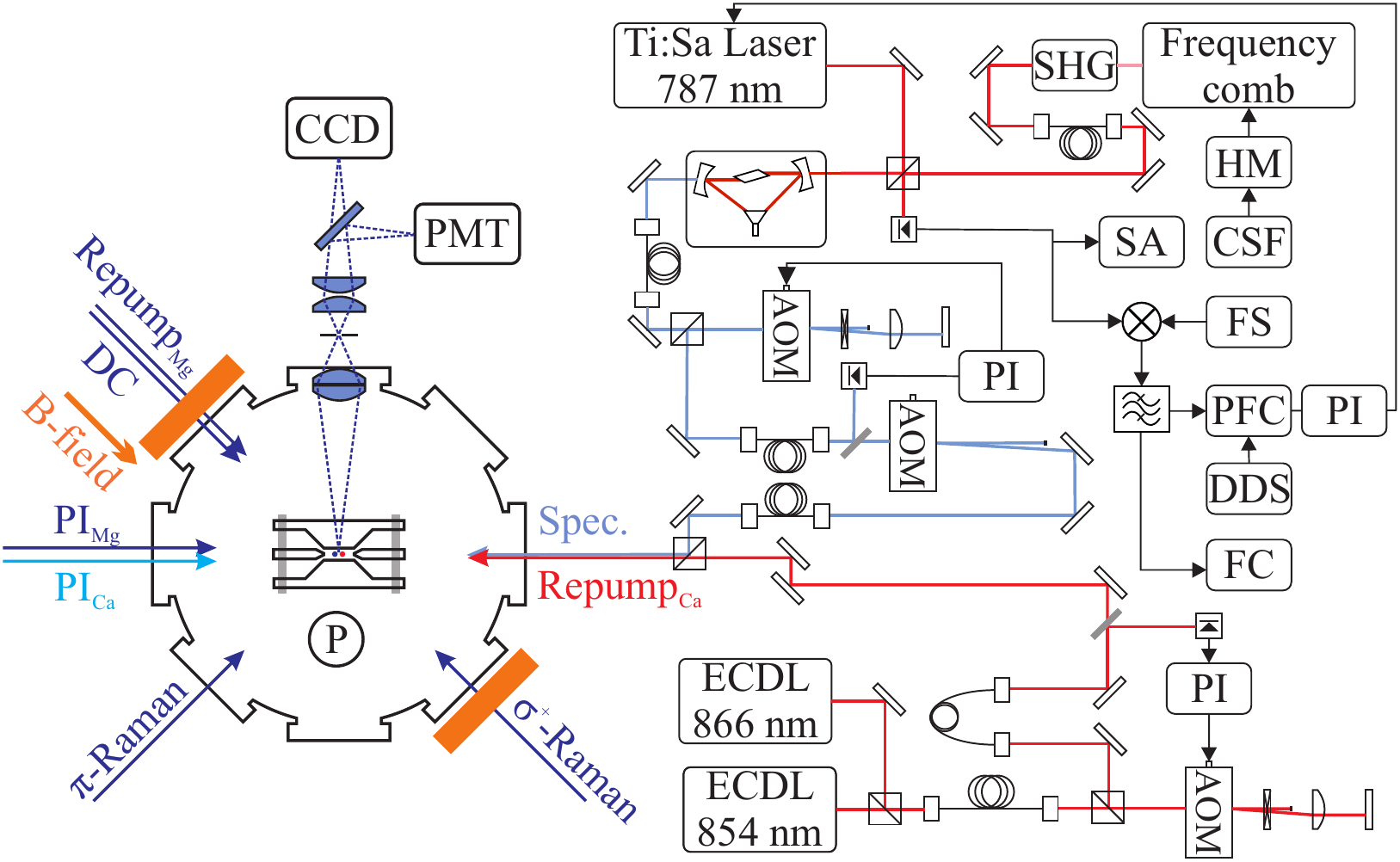}
\caption{Simplified experimental setup. The left part of this figure indicates the configuration of all optical beams and bias magnetic field with respect to the ions in the trap. The right part shows the optical setup for the spectroscopy beam and the calcium repump beam. For details, see text. All RF sources are referenced to a 10-MHz signal from the H-maser (HM). AOM: acousto-optical modulator; CCD: electron-multiplication charge coupled device; CSF: caesium fountain clock; DC: Doppler cooling beam; ECDL: external-cavity diode laser; FC: frequency counter;
FS: frequency synthesizer; P: Parabolic mirror; PFC: phase and frequency comparator; PI: proportional-integral controller; PI$_\mathrm{Mg/Ca}$: photo-ionization beam for Mg/Ca; PMT: photomultiplier tube; SA: spectrum analyzer; SHG: second harmonic generation.}
\label{fig:tisa}
\end{figure}

Spectroscopy is performed using a cw single-mode Ti:Sa laser (Sirah, Matisse TS) which is frequency doubled in an enhancement cavity (Spectra-Physics, WaveTrain). The frequency of the laser is locked to an erbium-fiber-laser-based frequency comb. The comb is stabilized in its offset and repetition frequencies to a H-maser frequency which is calibrated by a caesium fountain at PTB (German National Metrology Institute) as shown in figure~\ref{fig:tisa}. A beat signal between the spectroscopy laser and the nearest comb tooth of a narrow-band frequency-doubled output of the frequency comb is detected with a fast photo diode. The nearest comb tooth is identified by measuring the frequency of the spectroscopy laser with a wavemeter (High Finesse, model WS-7).

\begin{figure}[h]
\centering
\includegraphics[width=\linewidth]{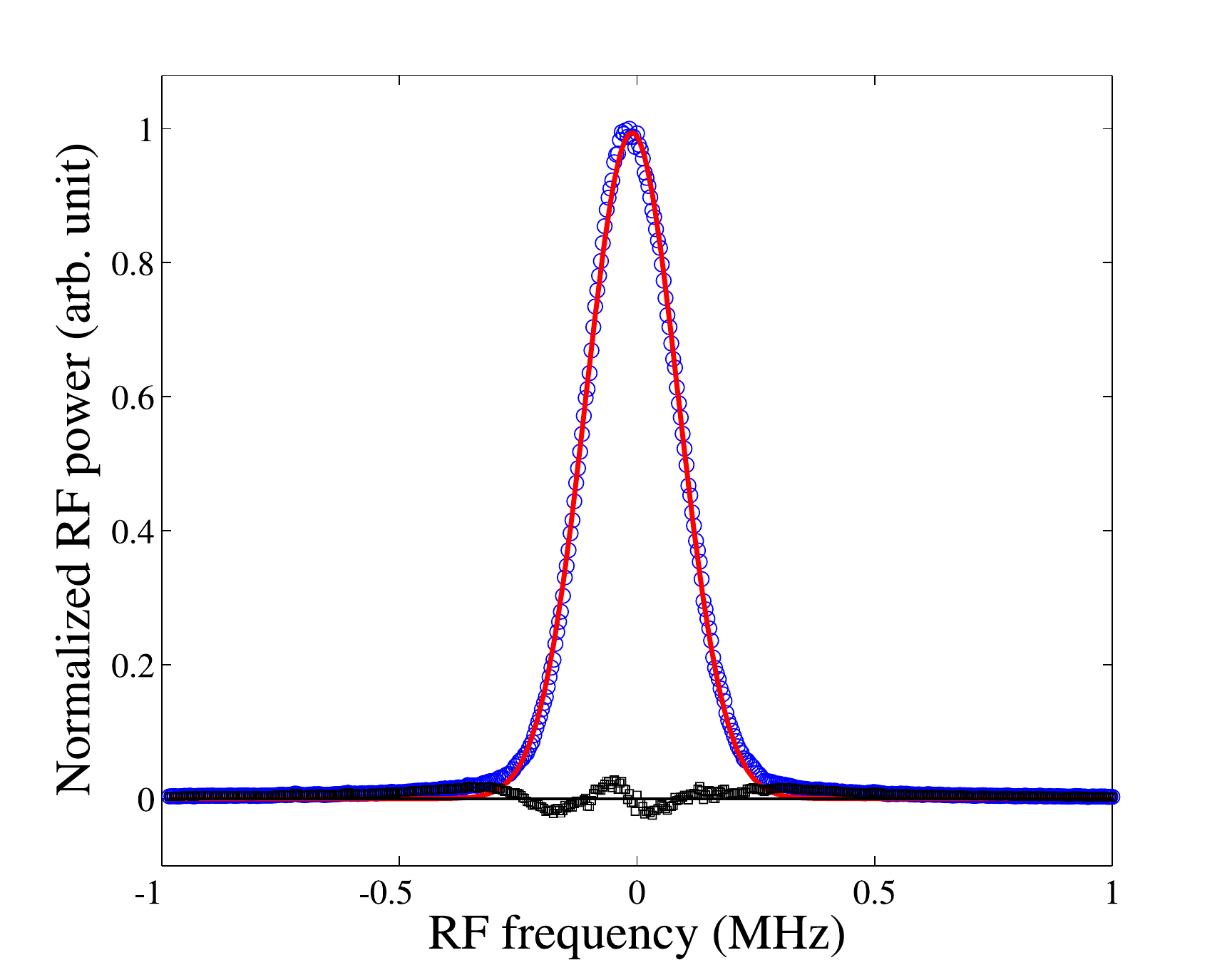}
\caption{In-loop beat signal of the Ti:Sa laser and the nearest frequency comb tooth. The data recorded by the spectrum analyzer with 10\,kHz resolution bandwidth (blue circles) is fit with a Gaussian profile (red curve) that gives an upper bound of 260\,kHz for the linewidth of the spectroscopy laser. The black squares present the residuals of the fit.}
\label{fig:beat}
\end{figure}

\begin{figure}[h]
\centering
\includegraphics[width=\linewidth]{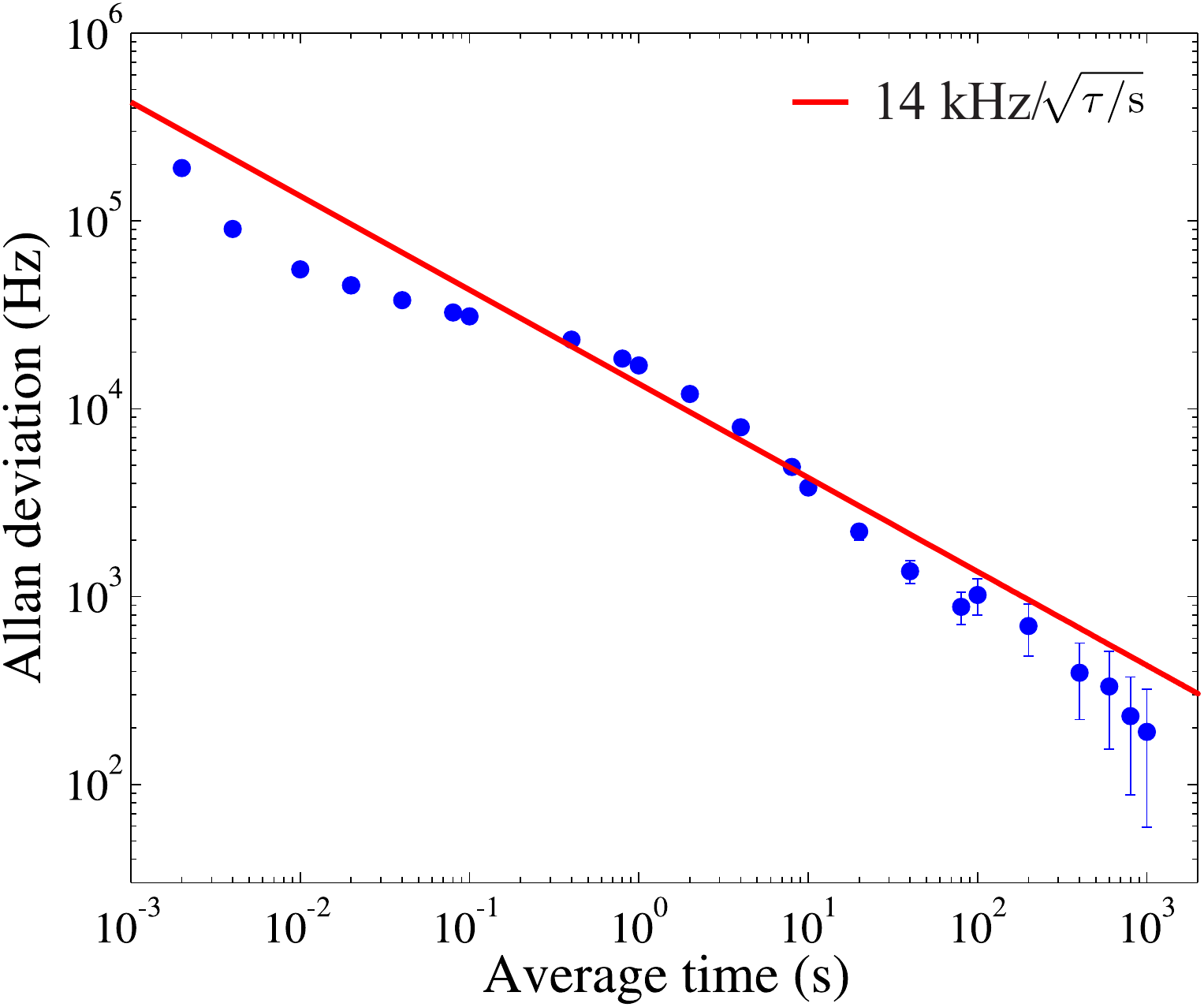}
\caption{Allan deviation of the Ti:Sa laser frequency as measured by the frequency counter. The long term instability of the locked spectroscopy laser is 14\,kHz$/\sqrt{\tau/\textrm{s}}$ for integration times longer than 1\,s as derived from a fit to the data (red line). }
\label{fig:allan deviation of the laser}
\end{figure}

The RF signal is monitored by a spectrum analyzer (Rohde \& Schwarz, FSL3) and mixed down to 10 or 35\,MHz using an rf synthesizer. The band-pass filtered signal is used as the input signal of a self-build phase frequency comparator (PFC) operating at 10 or 35\,MHz, which produces an error signal that is tailored by a proportional-integral (PI) controller. The generated control signal is used to correct the frequency of the Ti:Sa laser by changing the length of the laser cavity with a fast piezo-electric actuator. By adjusting the frequency of the rf synthesizer, the frequency of the spectroscopy laser can be adjusted to the resonances of the different isotopes.
Figure\,\ref{fig:beat} shows the in-loop beat signal between the Ti:Sa laser and the frequency comb, indicating an upper bound for the linewidth of the spectroscopy laser below 260\,kHz.
The long-term frequency instability is determined by counting the mixed-down RF signal with a frequency counter (Kramer+Klische FXE). We derive a frequency instability of about 14\,kHz/$\sqrt{\tau/\textrm{s}}$ from an Allan deviation of these measurements as shown in figure~\ref{fig:allan deviation of the laser}. The linewidth and the frequency instability both fulfil the requirements for the anticipated resolution and accuracy of below 100\,kHz. The spectroscopy beam is intensity stabilized and frequency scanned by an acousto-optical modulator (AOM), while another AOM is used for switching.

The 866~nm and 854~nm repump beams depopulating the two $D$-states are generated by two external-cavity diode lasers. The combined beam consisting of the two repumpers is intensity stabilized by an AOM which is also used for fast switching. The left part of Figure \ref{fig:tisa} shows the direction of the beams together with the magnetic bias field. It also contains the Doppler cooling, Raman, and repump beams for controlling the $^{25}$Mg$^+$ ion as described in detail e.g.\ in \cite{Wan2015}.

\section{\label{isotope shift}Isotope shift}
Isotope shifts in atomic transition frequencies originate from two effects: (\textit{i}) a change in the size of the nucleus and a corresponding change of potential experienced by electrons having a finite probability density inside the nucleus (field shift), and (\textit{ii}) a change of the center of mass motion of the nucleus (mass shift). The mass effect is more important for light nuclei due to the relatively large mass differences, while the field shift is proportional to the electron density inside the nucleus, which increases roughly with $Z^2$ and is the dominant contribution in elements with large atomic number $Z$ \cite{Breit1958}. The isotope shift is defined as $\delta\nu_{i}^{A,A'} = \nu_i^{A'} - \nu_i^{A}$, where $\nu_i^{A'}$ and $\nu_i^{A}$ are the frequencies of transition $i$ for isotopes with mass $m_{A}$ and $m_{A'}$, respectively.
The mass shift contribution to the isotope shift of a one-electron atom can be calculated by replacing the electron mass $m_e$ by the reduced mass of the system and leads to
\begin{equation}
\label{eq:ReducedMassIS}
\delta\nu_{i,\mathrm{NMS}}^{A,A'} = m_e \nu_i^{A} \times \frac{m_{A'}-m_{A}}{m_{A}\cdot (m_{A'}+m_e)}.
\end{equation}
It is called the normal mass shift (NMS) and the prefactor of the mass scaling  $\knms{i,} = m_e \nu_i^{A}$ is the normal mass shift constant. In a multi-electron system, the nuclear recoil depends on the sum of all electron momenta and therefore the mass polarization term or specific mass shift (SMS) can have a considerable contribution with the same mass-scaling. The corresponding constant \ksms{i,}\ is notoriously difficult to calculate since it includes all electron momenta.
The difference in the transition frequency, $\delta\nu_i^{A,A'}$, between isotopes with mass~$m_A$ and~$m_{A'}$ can in total be expressed as
\begin{equation}
\label{eq:IS}
\delta \nu_i^{A,A'} = \kms{i} \frac{m_{A'}-m_{A}}{m_{A}\cdot m_{A'}} + \fs{i} \, \drsq^{A, A'} \,,
\end{equation}
where $\kms{i} = \knms{i,} + \ksms{i,}$ is the (total) mass shift constant, \fs{i}\ is the field shift constant, and $\drsq^{A, A'}$ is the corresponding change in the mean-square nuclear charge radius of the two isotopes. Here we simplified the mass-dependence term by the approximated dependence as it is usally used in the literature. Neglecting the additional electron mass in the denominator of Eq.\,(\ref{eq:ReducedMassIS}) leads to a change only of the order of $(m_e/m_A)^2$.
To extract nuclear charge radii from isotope shifts, one needs reliable numbers for the mass shift and field shift constants, which can be obtained only from semi-empirical approaches or from ab-initio calculations. Reasonable agreement is usually obtained between these techniques. One of the most important procedures in this respect is the King plot \cite{King1984}, of which two versions exist that have different applications. Both will be used below when analyzing the data. \\
The general approach is to multiply both sides of Eq.\,(\ref{eq:IS}) with the inverse mass factor $\mu=m_{A}\cdot m_{A'}/ \left( m_{A'}-m_{A}\right)$ and obtain
\begin{equation}
\label{eq:modIS}
\mu \, \delta \nu_i^{A,A'} = \kms{i}  +  \fs{i} \, \mu \, \drsq^{A, A'} \,.
\end{equation}
The relation allows one to eliminate the unknown nuclear charge radii if measurements are performed in two different transitions $i$ and $j$ for the same isotopes
\begin{equation}
\label{eq:TransKingPlot}
\mu \, \delta \nu_i^{A,A'} = \kms{i} - \frac{\fs{i}}{\fs{j}}\kms{j} + \frac{\fs{i}}{\fs{j}} \, \mu \,\delta \nu_j^{A,A'} \,.
\end{equation}
This is a linear relation between the so-called modified isotope shifts $\mu \,\delta \nu^{A,A'}$ in the two transitions $i$, $j$ and can be used to extract the respective ratio of the field shift constants or the relation of the mass shift constants.
Alternatively, Eq.\,(\ref{eq:modIS}) can be used with data of a single transition to directly obtain field shift and mass shift constants. This requires mean-square charge radii or their respective changes $\drsq^{A, A'}$ for a subset of at least 3 isotopes from other sources. Usually, radii from elastic electron scattering or X-ray transitions in muonic atoms are used, see e.g., Ref. \cite{Fricke2004}. By plotting the modified isotope shift versus the modified changes in the rms charge radii $\mu \, \drsq^{A, A'}$, a linear regression delivers the field shift constant as the slope and the mass shift constant as the interception with the $y$-axis. A multidimensional regression can be used if information from several transitions is available.

Applying the extracted mass shift and field shift constants, charge radii of other, especially short-lived isotopes can be obtained. However, the accuracy of the extracted nuclear charge radii with these procedures is often hampered by the insufficient accuracy of the charge radii data from external sources or the limited number of stable isotopes. Odd-$Z$ elements for example have only one or two stable isotopes, which renders the usage of the King-plot procedure impossible.
High-precision isotope shift data with accuracy better than 1\,MHz has rarely been obtained beyond the lightest elements hydrogen \cite{Schmidt-Kaler1995}, helium \cite{Zhao1991,Marin1994,Shiner1995,Wang2004,Mueller2007,Pastor2012}, lithium \cite{Riis1994,Ewald2004,Sanchez2006,Noertershaeuser2011,Sansonetti2011,Brown2013} and beryllium \cite{Noertershaeuser2009,Krieger2012,Noertershaeuser2015} where they are used to either extract nuclear charge radii of stable and short-lived isotopes or to test many-body non-relativstic quantum electrodynamics calculations (NR-QED). Similarly accurate data does so far only exist for the D1 and D2 lines in magnesium \cite{Herrmann2009,Batteiger2009} and the D1 line in calcium \cite{Wan2014,Gebert2015}, all obtained by laser spectroscopy in Paul traps. Small differences in mass corrected isotope shifts between transitions of the same fine structure multiplet provide information on subtle (relativistic) effects of the electronic wavefunction. A $J$-dependence of the field shift
constant has already been reported in the $6s \rightarrow 6p$ doublet in Ba II, where a field shift difference of 2.5(3)\,\% was observed \cite{Wendt1984}.
In lighter isotopes these effects, caused by different contributions of the smaller component of the Dirac wavefunction, are expected to be much smaller and were so far not reported. Even in the most precise ab-initio calculations up to Be, field shift factors are assumed to be equal for both transitions of the respective doublets, whereas a small relativistic mass-dependent change was included. In the case of Mg, the transition frequencies were measured only in two isotopes and, thus, a King-plot analysis could not be performed. Here, we report on high-precision calcium isotope shift measurements in the D2 line for $^{40,42,44,48}$Ca from which we extract differences in the probability density of the electrons at the nucleus and the mass dependence of the difference between the isotope shifts in the two
transitions $i,j$ of a fine structure doublet, the so-called splitting isotope shift (SIS) 
\begin{equation}
 \delta\nu^{A,A'}_{\mathrm{SIS}} = \delta \nu^{A,A'}_{i} - \delta \nu^{A,A'}_{j},
\end{equation}
which can also be understood as the change of the fine structure splitting between the respective isotopes.

\section{\label{sec:result} Experimental results}

\subsection{Transition frequencies}
\begin{figure}[h]
\centering

\includegraphics[width=\linewidth]{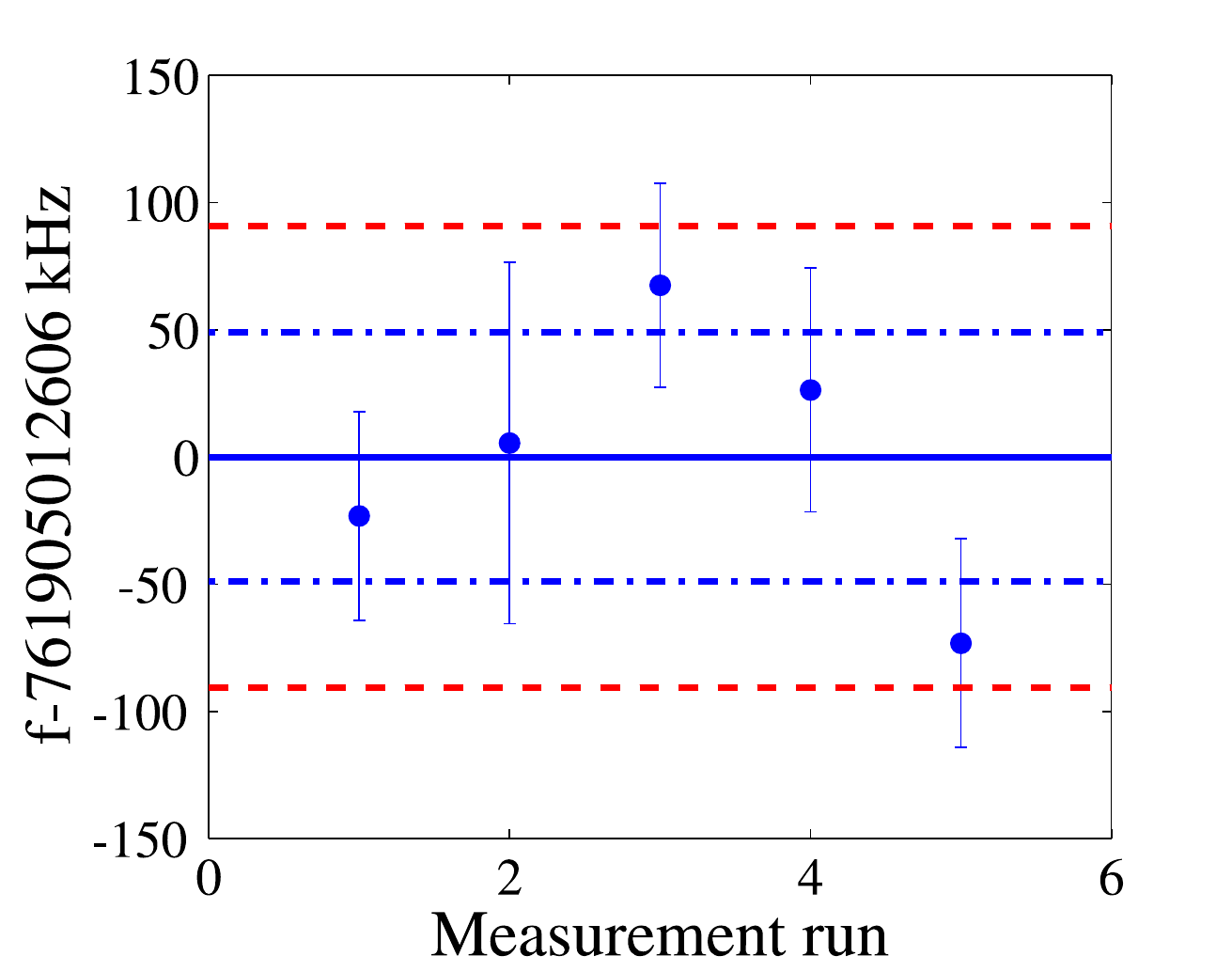}
\caption{Results of five independent measurements of the absolute transition frequency in the D2 line of \Ca{40}. The blue solid line is the weighted average frequency. The blue dashed-dotted lines represent the statistical uncertainty and the red dashed lines the combined uncertainty corresponding to a confidence interval of 68.3\,\%. The error bar assigned to each measurement point indicates the statistical uncertainty.}
\label{fig:afd2}
\end{figure}

The absolute frequency of the D2 transition of $^{40}$Ca$^{+}$ is measured using the photon recoil spectroscopy technique with an accuracy of better than 100\,kHz. Table\,\ref{tab:absolute frequency} lists this new value together with previously measured transitions using the same technique \cite{Wan2014,Gebert2015} and literature values. As for the previous measurements, the main systematic shifts of the D2 transition frequency include the lineshape shift, AC-Stark shift, Zeeman shift, and AOM-envelope shift. We evaluate the shifts following the procedures outlined in reference \cite{Wan2014} and adopt the results for the Zeeman and Stark shift measurements.
The frequency shift estimation is listed in Table\,\ref{tab:af d2}. The absolute frequency of the $^{40}$Ca$^{+}$ D2 transition is 761\,905\,012\,606\,(91)\,kHz obtained as the weighted average of the five measurements shown in Fig.\,\ref{fig:afd2}.

\begin{table}[tbh]
\centering
\caption{Absolute frequency of the \dsohdpth, \dsohdpoh, and \ddthdpoh
transition of $^{40}$Ca$^{+}$.}
\label{tab:absolute frequency}
\begin{tabular}{ldc}
\hline
\hline
\multicolumn{1}{c}{Transition} & \multicolumn{1}{c}{\textrm{Frequency (MHz)}} & \multicolumn{1}{c}{Ref.}\\
\hline
\dsohdpth
&  761\,905\,012.606\;(91)  & this work\\
&  761\,905\,012.7\;(5)  & \cite{Wolf2009}\\
\dsohdpoh
& 755\, 222\,765.896\;(88) & \cite{Wan2014}\\
& 755\, 222\,766.2\;(17) & \cite{Wolf2008} \\
\ddthdpoh
& 346\,000\,234.867\;(96)  & \cite{Gebert2015} \\
 \hline
 \hline
\end{tabular}
\newline
\end{table}

\begin{table} []
\centering
\caption{Uncertainty estimation of the absolute D2 transition frequency in $^{40}$Ca$^{+}$. All values are in kHz.}
\label{tab:af d2}
\begin{tabular}{lcc}
\hline
\hline
Systematic effect & Shift & Uncertainty \\
\hline
Zeeman (static magnetic field) & -8 & 59\\
AC Stark (spectroscopy laser) & 60 & 44\\
Lineshape (detection scheme) & 152 & 20\\
Spectroscopy laser lock & 0& 0.6\\
Statistics &0& 49\\
Total &204 & 91\\
\hline\hline
\end{tabular}
\end{table}

\subsection{\label{analysis}Isotope shift}

We analyse the data to extract the different contributions to the isotope shift and to establish a self-consistent set of observables that can be used to provide benchmarks for atomic structure calculations of this 19-electron system.
The isotope shifts are evaluated with respect to the most abundant isotope \Ca{40}. The transition frequencies of the reference and the isotope \Ca{A} are measured interleaved on one day to cancel all common systematic shifts. The resulting isotope shifts are listed in Table~\ref{tab:isotope shifts} together with the isotopic shift of the D1 transition obtained in a previous measurement \cite{Gebert2015}, and their difference. The latter, the so-called splitting isotope shift $\delta\nu_\mathrm{SIS}$ will be discussed in Sec.~\ref{sec:sis}.
\begin{table}[b]
\caption{Measured isotope shifts in the D1 and D2 transitions referenced to \Ca{40}. All values are given in MHz. The last column provides the difference between the isotope shifts in the two lines of the fine structure doublet, which is commonly known as the splitting isotope shift $\delta\nu_\mathrm{SIS}^{A,40}$. The uncertainties are dominated by statistics.}
\label{tab:isotope shifts}
\centering
\begin{tabular}{cbbb}
\hline\hline
A & \multicolumn{1}{c}{$\delta\nu^{A,40}_{\mathrm{D1}}$} &\multicolumn{1}{c}{$\delta\nu^{A,40}_{\mathrm{D2}}$} & \multicolumn{1}{c}{$\delta\nu_\mathrm{SIS}^{A,40}$} \\
\hline
42 &  425.706(94) &  425.932(71) & 0.226(118)\\
44 &  849.534(74) &  850.231(65) & 0.697(98)\\
48 & 1705.389(60) & 1707.945(67) & 2.556(90)\\
\hline\hline
\end{tabular}
\end{table}

\begin{figure}[h]
\centering
\includegraphics[width=\linewidth]{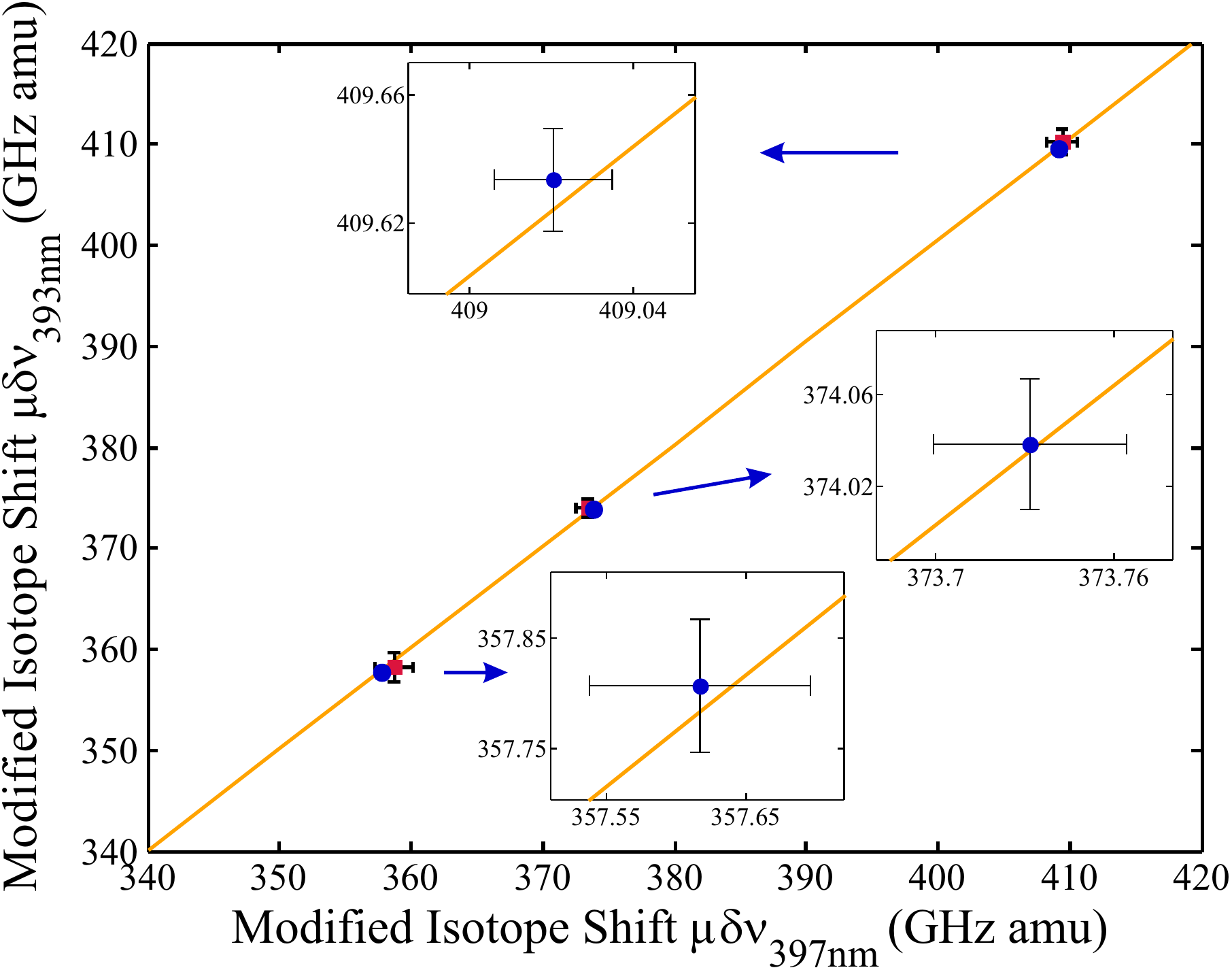}
\caption{King plot of the D2 versus the D1 transition. Plotted are the modified isotope shifts $\mu\,\delta\nu^{40,A}_{i}$ in the D2 against the D1 transition. The red dots represent the formerly most accurate measurements \cite{Gorges2015} and the blue dots are the measurements presented here, with uncertainties smaller than the symbol. The line is the result of a linear regression of Eq.\,(\ref{eq:TransKingPlot}) taking uncertainties in both axis into account (for details see text). The insets show the relevant ranges around the data points enlarged by more than two orders of magnitude to illustrate the quality of the fit.}
\label{fig:King393397}
\end{figure}

We performed a King plot analysis of the D1 versus the D2 transition according to Eq.\,(\ref{eq:TransKingPlot}). The result is shown in Fig.\,\ref{fig:King393397} and demonstrates the high quality of the data indicating that at this level of accuracy second-order mass polarization terms can still be neglected. The slope of the line provides the ratio of the field shift constants $f:=\fs{D2}/\fs{D1}$ in the two transitions while a relation between mass and field shift is obtained from the intersection with the $y$-axis as $k:=K_{\mathrm{D2}}-K_{\mathrm{D1}} F_{\mathrm{D2}}/F_{\mathrm{D1}}$. We used two fitting routines that are able to take the measurement uncertainties in $x$ and $y$ direction into account to determine the parameters $f$ and $k$ with corresponding uncertainties: we employed the algorithm described by York et al.\ \cite{York2004} and performed a Monte Carlo analysis \cite{Gebert2015}. The results of both methods are listed in Table~\ref{tab:res2DKing} and are fully consistent. A standard linear
regression performed with MATLAB gives consistent values for the fit parameters, but differs in the assigned uncertainty, since it neglects the $x$ uncertainty of the experimental data.
\begin{table}[tbh]
\caption{Ratio of field shift factors $f:=\fs{\mathrm{D2}}/\fs{\mathrm{D1}}$, $f_{1}:=\fs{\mathrm{DP}}/\fs{\mathrm{D1}}$, $f_{2}:=\fs{\mathrm{DP}}/\fs{\mathrm{D2}}$ and difference $k:=K_{\mathrm{D2}}-f \cdot K_{\mathrm{D1}}$, $k_{1}:=K_{\mathrm{DP}}-f_{1} \cdot \kms{\mathrm{D1}}$,
$k_{2}:=K_{\mathrm{DP}}-f_{2} \cdot \kms{\mathrm{D2}}$ in \GHzamu\ as obtained from the King plots of the D1, D2 and \ddthdpoh (DP) transitions.}
\label{tab:res2DKing}
\centering
\begin{tabular}{cbbb}
\hline\hline
  & \multicolumn{1}{c}{\textrm{York et al.}}
  & \multicolumn{1}{c}{\textrm{Monte Carlo}}
  & \multicolumn{1}{c}{\textrm{std. fit}}\\
\hline
$f$ &      1.0085(11) &  1.0085(12) &  1.0083(6)  \\
$k$ &     -2.881(472) & -2.873(473) & -2.787(212) \\
$f_{1}$ & -0.3110(10) & -0.3114(10) & -0.3116(15) \\
$k_{1}$ & -1862.9(4)  & -1862.8(4)  & -1862.7(6)  \\
$f_{2}$ & -0.3084(10) & -0.3088(10) & -0.3090(17) \\
$k_{2}$ & -1863.8(4)  & -1863.6(4)  & -1863.6(7)  \\

\hline\hline
\end{tabular}
\end{table}
From $f$ we can clearly conclude that the field shift in the D2 transition is by 0.85(12)\,\% larger than in the D1 transition.
The size of this difference comes as a surprise compared to a simple estimate using the hydrogenic approach of 0.52\,\% which is expected to provide an upper bound and will be discussed in Sec.~\ref{sec:theory}.

Absolute values for the field-shift and mass-shift constant are required to extract nuclear properties. This is not possible solely based on spectroscopic data without additional information on the finite nuclear size effect. Fortunately, there is plenty of data for the stable calcium isotopes not only for the mean square charge radius but also for form factors and their isotopic change. The most intriguing point in the calcium isotope chain is the fact that the two doubly magic isotopes \Ca{40,48} have practically identical mean-square charge radii. Form factor measurements indicate that this is due to the fact that charge is being transferred from the center and the skin of the nucleus towards the surface region where the nuclear density dropped to about half the saturation value, resulting in identical mean-square charge radii for both isotopes \cite{Rebel1979,Emrich1983,Otten1989}. Since there is clear evidence for this from non-optical data and it is also confirmed with high accuracy from optical data, we use this particularity of the isotope chain to separate mass and field shift. Below, we will present the result of a full analysis with all uncertainties included. First we assume that $\drsq^{48,40}=0$ in order to explore the limits inherent in our measurement uncertainty rather than being limited by the uncertainty of the nuclear size correction. With this assumption, the isotope shift between the doubly magic isotopes arises entirely by the mass shift, which is $K_\mathrm{D1}=409.020(14)[304]\,$\GHzamu and $K_\mathrm{D2}=409.633(16)[307]\,$\GHzamu for the D1 and D2 transitions, respectively. The parentheses represent the uncertainty excluding any uncertainty of $\drsq^{40,48}$, while the value in square brackets indicates the change of $K_i$ for $\drsq^{40,48}=-0.0045$\,fm$^2$ taken into account. The significant deviation of the ratio $K_\mathrm{D2}/K_\mathrm{D1}=1.00150\,(5)$ from one is caused by relativistic effects. Please note that the uncertainty from $\drsq^{40,48}$ given in the square brackets largely cancels in this ratio since it changes both mass shift constants by the same (small) amount.

For a full analysis we include the newly measured transition and extend the analysis performed in \cite{Gebert2015}. From this, we extract field and mass shift constants for the D2 line and improve the uncertainties of the constants in the D1 transition. The result of this analysis is displayed in Table~\ref{tab:parameters}.
\begin{table}[h]
\caption{Parameters of a three-dimensional King plot seeded with values of $\drsq^{A,40}$ taken from \cite{Wohlfahrt_muonic_1978}. The units for the field shift constants $\fs{i}$ and mass shift constants $\kms{i}$ and the changes in mean square nuclear charge radii $\drsq^{j,40}$ are \MHzfmsq, \GHzamu\ and \fmsq, respectively. \label{tab:parameters}}
\begin{center}
\begin{tabular}{lbb}
\hline\hline
Param. & \multicolumn{1}{l}{Previous \cite{Gebert2015}} & \multicolumn{1}{l}{This work} \\
\hline
$F_{\mathrm{D1}}$      &  -281.8(7.0) &  -281.8(6.9) \\
$K_{\mathrm{D1}}$      &   408.73(40) &   408.73(40) \\
$F_{\mathrm{D2}}$      &              &  -284.7(8.2) \\
$K_{\mathrm{D2}}$      &              &   409.35(42) \\
$F_{\mathrm{DP}}$ &    87.7(2.2) &    87.6(2.2) \\
$K_{\mathrm{DP}}$ & -1990.9(1.4) & -1990.0(1.2) \\
$\drsq^{42,40}$ &  0.2160(49) &   0.2160(49) \\
$\drsq^{44,40}$ &  0.2824(65) &   0.2824(64) \\
$\drsq^{48,40}$ & -0.0045(60) &  -0.0045(59) \\
\hline\hline
\end{tabular}
\end{center}
\end{table}

\section{Splitting isotope shift}\label{sec:sis}

\begin{figure}[tbp]
\centering
\includegraphics[width=\linewidth]{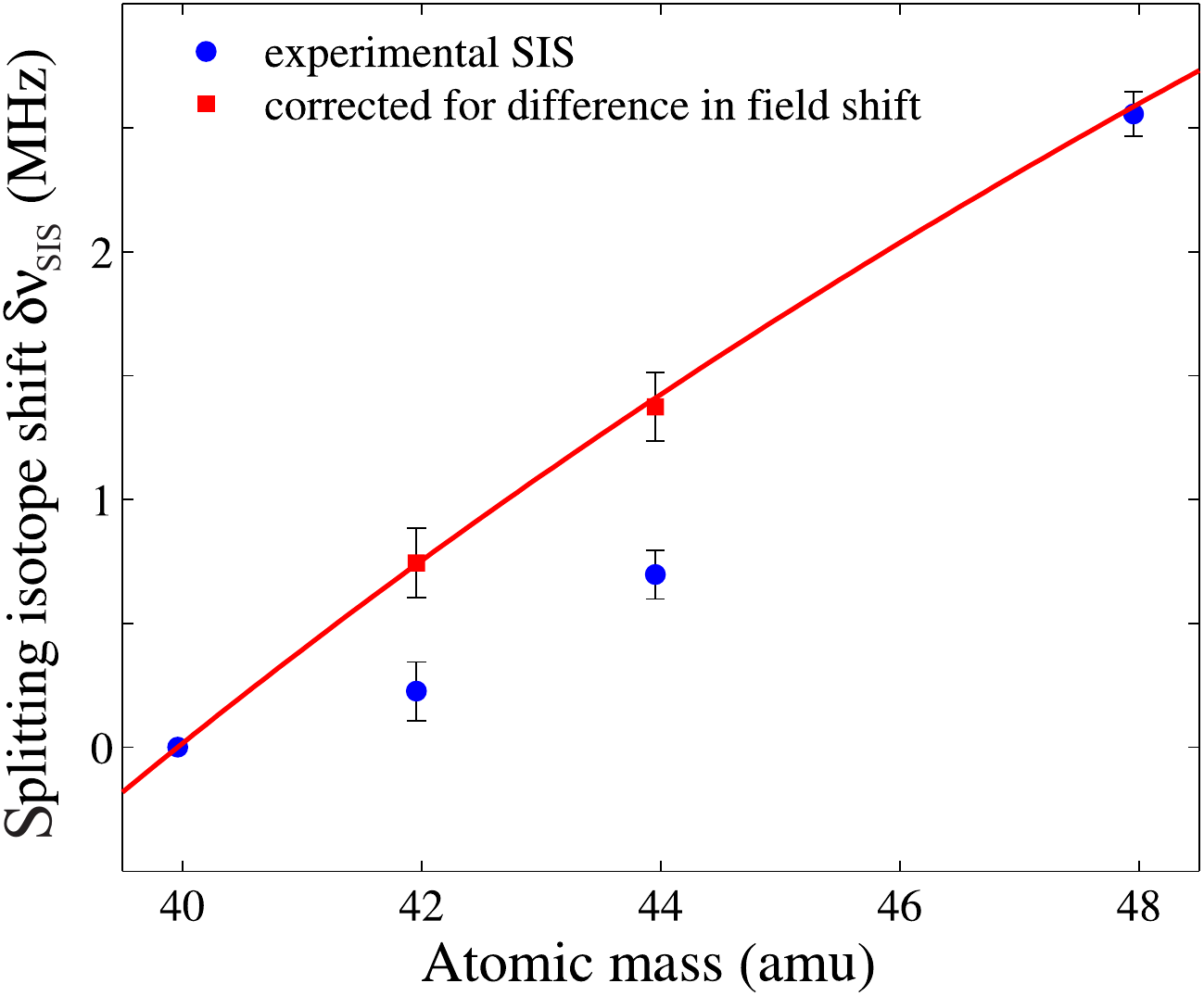}
\caption{Splitting isotope shift as extracted from the measurements in the D1 and the D2 transitions. The blue circles are the experimentally determined splitting isotope shifts. The red line represents the mass dependence as expected from the measured $\delta\nu_\mathrm{SIS}^{40,48}$ assuming that $\drsq^{40,48}=0$ and neglecting the field shift contribution. The red squares are the SIS of \Ca{42} and \Ca{44} after correction for the remaining field shift in the SIS.}
\label{fig:SIS_plot}
\end{figure}

The splitting isotope shift (SIS), i.e.\ the change of the fine structure splitting between isotopes, has been recently the subject of investigations in He \cite{Pachucki2011}, Li \cite{Sansonetti2011,Brown2013} and Be \cite{Noertershaeuser2015}. It is known to have in first order a mass dependence linear in $1/\mu=1/m_A - 1/m_{A'}$. For light isotopes, the SIS is nearly independent of both, QED and nuclear volume effects and has therefore served as an important consistency check for theory and experiment \cite{Puchalski2013}. From our previous analysis of the isotope shifts, we can conclude that in calcium a small contribution of the finite nuclear size effect is still inherent in the SIS since the field shift coefficients for both transitions are slightly different. Again, we can utilize the identical mean-square charge radii of \Ca{40,48}, extract the mass polarization factor $\kms{\mathrm{SIS}}=-613\,(21)\, \mathrm{MHz \cdot amu}$ and plot the mass dependence of the SIS (red solid line in Fig.\,\ref{fig:SIS_plot}). The experimental SIS values for \Ca{42,44} clearly deviate from this prediction for light isotopes. However, the typical mass-dependence is restored to very high accuracy if the SIS is corrected for the difference in the field shift contribution according to
\begin{equation}\label{eq:fscorr}
\delta \nu_\mathrm{SIS,fs-corr} = \delta \nu_\mathrm{SIS,exp} - \left( \frac{F_{\rm D2}}{F_{\rm D1}}-1 \right) \cdot F_\mathrm{D1} \cdot \delta \left \langle r^2_c \right \rangle
\end{equation}
using the field shift ratio $f$ as determined above, the field shift factor $F_\mathrm{D1}=-284.7(8.2)$\,MHz/fm$^2$ and the known change in the mean-square charge radii according to Table\,\ref{tab:parameters}. 
The obvious discrepancy between the expected mass dependence of the first-order mass polarization term and the experimentally observed SIS for \Ca{42,44} can be perceived as the first detection of the field shift in a fine structure transition ($\dpoh \rightarrow \dpth$)
of a light ion.

In conclusion, the experiment provided high accuracy data that can now be used to guide and benchmark theoretical mass shift and field shift calculations.
In the following section we present improved calculations of the field shift in these transitions.

\section{Theoretical background}
\label{sec:theory}

Theoretical analysis of the isotope shift in heavy atoms is generally a rather complicated task which requires a proper account
of relativistic, many-body and even quantum electrodynamics (QED) effects. Since the detailed evaluation of these effects is beyond the scope of the present paper, we just briefly recall main theoretical approaches and their predictions here. We will focus again on the field shift constants $F_i$ that characterize the finite nuclear size contribution to the difference in transition frequencies
$\delta\nu_i^{A,A'}$, see Eq.~(\ref{eq:IS}). Our experiment has shown that the field shift in the D2 (\dsohdpth) transition is clearly larger than that in the D1 (\dsohdpoh) channel. This $J$-dependence of the constant $F_i$ can be understood from the different behaviour of the Dirac wave functions of the $4p_{\nicefrac{1}{2}}$ and $4p_{\nicefrac{3}{2}}$ states at the coordinate origin. Namely, while $\psi_{4p_{\nicefrac{3}{2}}}(r)$ vanishes at $r  = 0$, the $\psi_{4p_{\nicefrac{1}{2}}}(0)$ is nonzero due to the contribution of its small (relativistic) component with the orbital angular momentum $\ell=0$.  Hence the reduction of the electron density at the nucleus, being the origin of the field shift, is larger for the \dsohdpth transition.

Having discussed the (qualitative) reason for the difference of the $D1$ and $D2$ field shifts, we will compute now the ratio
$f = F_{\mathrm{D2}}/F_{\mathrm{D1}}$. As a first approximation, we employ the known formulas for the finite--nuclear--size correction for the
$n = 4$ \textit{hydrogenic} states from Ref.~\cite{MoT12} to find:
\begin{equation}
\label{eq:hydronic_ratio}
f_{\rm hydr} = 1 + (Z \alpha)^2 \frac{15}{64} + O((Z \alpha)^2) \, .
\end{equation}
This expression yields $f_{\rm hydr} = 1.0050$ for the nuclear charge $Z = 20$, which is in good agreement with the result
$f_{\rm hydr,num} = 1.0051$ of a direct numerical evaluation of hydrogenic field shifts.

Within the na\"ive hydrogenic approach one can also estimate the contribution of the QED effects to the ratio $F_{\mathrm{D2}}/F_{\mathrm{D1}}$
of the field shift constants. Namely, by using results for nuclear-size correction to the Lamb shift of one-electron atoms,
Ref.~\cite{Yer11}, we estimate:
\begin{equation}
\label{eq:QED_correction_hydrogenic}
\delta f^{\rm QED}_{\rm hydr} = (Z \alpha)^2 \,\frac{15}{64}\times
\frac{\alpha}{\pi}\,[1 + O(Z \alpha)] \, .
\end{equation}
This implies that the QED effects are negligible for the $f$ ratio in the Ca$^+$ ion, $\delta f^{\rm QED}_{\rm hydr} \ll f_{\rm hydr}$.
Therefore, any further corrections to the $f_{\rm hydr}$ may arise only from the electron-electron ($e$--$e$) interactions which were
neglected in the hydrogenic model. These interactions should \textit{reduce} the $f_{\rm hydr}$ as can be expected, for example,
from Eq.~(\ref{eq:hydronic_ratio}) in which we can decrease the charge $Z$ to account for the screening of the nucleus by core
electrons.

\begin{table}[h]
\caption{Transition field shift constants and ratios obtained in the CCSD(T) and CI+MBPT calculations and comparison with experimental values. We use the notation defined in Table~\ref{tab:res2DKing}. $F$ is in MHz/fm$^2$, $f$ dimensionless. \label{tab:comparison_theory_exp}}
\begin{center}
\begin{tabular}{lbbbb}
\hline\hline
Param. & \multicolumn{1}{c}{Experiment} & \multicolumn{1}{c}{CCSD(T)} & \multicolumn{1}{c}{CI+MBPT}\\
\hline
$F_{\mathrm{D1}}$      &  -281.8(7.0) &  -279.0(6.0) & -288.6(1.2)\\
$F_{\mathrm{D2}}$      &  -284.7(8.2) &  -280.3(6.0) & -289.0(1.2)\\
$f$ & 1.0085(12) & 1.0048 & 1.0014(4)\\
$F$ 
&    87.7(2.2) & 103(10) & 90.3(1.0)\\
$f_2$ 
& -0.3088(10) & -0.367 & -0.312(5)\\
\hline\hline
\end{tabular}
\end{center}
\end{table}

\begin{table}
\caption{Theoretical predictions for the ratio $f = F_{\mathrm{D2}}/F_{\mathrm{D1}}$ of the field-shift constants for the Ca$^+$ ion. The theoretical results are compared, moreover, with the present experimental value of $f$.
\label{tab:theo}
}
\begin{tabular}{ldl}
\hline\hline  \\[-0.3cm]
Theoretical model & \multicolumn{1}{c}{$f$} & Refs. \\
\hline \\[-0.3cm]
Hydrogenic  & 1.0051 & This work \\
Dirac-Fock  & 1.0010 & This work \\
Dirac-Fock + Core Pol. & 1.0009 & This work \\
CCSD                   & 1.0029  & This work \\
CCSD(T)                & 1.0048  & This work \\
MBPT                   & 1.0011 & Ref.~\cite{SaJ01} \\
CI+MBPT & 1.0014\;(4) & This work\\
\hline \\[-0.3cm]
Experimental value &   1.0085\;(12) & This work\\
\hline\hline
\end{tabular}
\end{table}

In order to describe properly the many-electron contributions to the field-shift ratio $f$, one needs to apply theories more advanced than the ``screening'' hydrogenic model. Therefore, as a second approximation, we solved the DF equation
and found the finite nuclear size corrections to the Ca$^+$ energy levels $\Delta E_{\rm fns}$. The field-shift ratio is then simply evaluated as $f_{\rm DF} = \Delta E_{\rm fns}(D1) / \Delta E_{\rm fns}(D2)$ and is
1.0010 for Ca$^+$. This DF result includes the first order $e$--$e$ correlations and can be further improved by
taking into account the higher--order corrections. For example, to consider (partially) the second--order $e$--$e$ effects we
employed the DF equation with the effective core--polarization potential \cite{MiN88,NoS76,Yer16}. As seen from the third line of the
Table~\ref{tab:theo}, this leads to a further slight reduction of the $f$ ratio.

To further investigate the role of $e$--$e$ interactions, we employ a RCC theory
starting with the DF equation by considering singles and doubles excitation approximation (CCSD method) and accounting important
triple excitations in the CCSD method (referred to as CCSD(T) method) as described in Ref. \cite{Sahoo2010,Wansbeek2012}.
In this calculation, we consider a Dirac-Coulomb-Breit Hamiltonian and lower order QED corrections in the approximations described in Ref.~\cite{Sahoo2016}. We obtain the field shift constant for the D1 and D2 lines as $F_{\mathrm{D1}}=-284.0$\,MHz/fm$^2$ and $F_{\mathrm{D2}}=-284.8$\,MHz/fm$^2$ in the CCSD method and $F_{\mathrm{D1}}=-279.0$\,MHz/fm$^2$ and $F_{\mathrm{D2}}-280.3$\,MHz/fm$^2$ in the CCSD(T) method.  This corresponds to $f_{\rm CCSD}=1.0029$ and $f_{\rm CCSD(T)}=1.0048$ in the CCSD and CCSD(T) methods respectively. We compare the experimental results on the field shift constants and their ratios in Table~\ref{tab:comparison_theory_exp} with the ones obtained from the CCSD(T) method.
The ratio obtained in CCSD(T) is close to the hydrogenic value, but still considerably smaller than the experimental result.

As an alternative approach to the CCSD(T) method we have examined the correlation potential method~\cite{dzuba87jpb} and the combination of configuration interaction and many-body perturbation theory (CI+MBPT)~\cite{dzuba96pra,berengut06pra}. Both methods give very consistent results. In the first method we create a correlation potential $\hat\Sigma$ to second order in the residual Coulomb interaction. We include this potential in the DF Hamiltonian and solve to create ``Brueckner'' orbitals. The field shift is obtained by varying the nuclear radius, repeating the entire calculation, and extracting the dependence of the energy eigenvalue. This `finite field' method is similar to that previously applied to calculations of field shift in Ca$^+$~\cite{berengut03pra}. It includes certain chains of diagrams to all-orders, for example so-called random-phase approximation corrections are included. Our result for the D2/D1 ratio using Brueckner orbitals is $f = 1.0010$. This value is directly comparable to the third-order MBPT results of Safronova and Johnson~\cite{safronova01pra} who also obtain $f = 1.0010$.

In the second method we perform a configuration interaction (CI) calculation allowing for single and double excitations from the $3s$ and $3p$ hole states. Correlations with the frozen core ($K$ and $L$ shells) are included at second-order in MBPT by modifying the radial integrals of the CI calculation. This `particle-hole' CI+MBPT method is similar to that introduced in Hg$^+$~\cite{berengut16pra}. It improves upon the Brueckner calculation in that correlations with the core $3s$ and $3p$ shells are taken into account to all order (only double excitations are included, but triple excitations in this calculation are not important). Again we use the finite-field method to obtain the field shift. The final results presented in Tables 7 and 8 is the CI+MBPT method, with the difference from the Brueckner calculation taken as an indicative uncertainty. Note that QED effects are not included in these results. They can contribute up to 2~MHz/fm$^2$, to $F_\textrm{D1}$ and $F_\textrm{D2}$, but as discussed previously they have a much smaller effect on the ratio.

In Table~\ref{tab:theo} we summarize our theoretical predictions of the field--shift ratio $f$. As expected, the hydrogenic value provides the upper bound of the ratio $F_{\mathrm{D2}}/F_{\mathrm{D1}}$ while the electron correlations lead to a reduction of $f$. Moreover, one can see from the table that the experimental result for the relativistic correction to the ratio of field-shift constants, i.e. $1 - f_{\rm exp}$, is 70\,\% larger than all available calculations, including even $f_{\rm hydr}$. The discrepancy between $f_{\rm exp}$ and the best theoretical models is about 3$\sigma$ of the experimental uncertainty; its reason is still unclear and will require future study.

\section{\label{sec:summary}Summary}
We presented absolute frequency and isotope shift measurements of the D2 line in even calcium isotopes at the 100~kHz level. The high accuracy of the measurement was enabled through photon recoil spectroscopy on trapped and sympathetically cooled ions using a spectroscopy laser referenced via an optical frequency comb to a calibrated hydrogen maser at PTB. From a multi-dimensional King plot analysis including data from previous measurements on the D1 and \ddthdpoh transitions, we derived a slightly improved set of field and mass constants for these transitions, and changes in the nuclear charge radius between isotopes. In particular, the measurements revealed for the first time field shift contributions to the splitting isotope shift in a light ion. Using the experimental data as a benchmark, we have performed theoretical calculations of the field shift constants and their ratios by using a simple hydrogenic approach, by solving the Dirac-Fock equation, also including core polarization, by relativistic coupled-cluster calculations including up to triples excitation, and by combining configuration interaction and many-body perturbation theory. The individual field shift constants for the D1 and D2 line derived from the coupled-cluster and many-body perturbation theory calculation show satisfactory agreement with the experimental data. However, we experimentally found a surprisingly large ratio of the field shifts in the D2 and D1 fine structure duplet, $f_{\rm exp}$. The relativistic correction to this value, $1- f_{\rm exp}$, is about 70\,\% larger than the theoretical estimates, which is a 3 $\sigma$ of experimental uncertainty. The account of electron-electron interactions leads to a further reduction of the theoretical value $f_{\rm theo}$ and, hence, to an even larger discrepancy with the experiment. For example, based on the CCSD theory we found that the difference of the D1 and D2 field shifts of 0.18\,\% is considerably smaller than the 0.83\,\% found in the experiment. The origin of this unexpectedly large difference must be clarified by further theoretical and experimental studies.

\section*{Acknowledgement}
We acknowledge financial support from the German Federal
Ministry for Education and Research (BMBF) under contract
05P15RDFN1, the Helmholtz International Center for FAIR (HIC for FAIR)
within the LOEWE program by the State of Hesse, the State of Lower-Saxony, Hannover, Germany and DFG through grants SCHM2678/3-1 and CRC~1227 DQ-\textit{mat}, project B05. SK received support from HGS-Hire. BKS acknowledges use of the Vikram-100 HPC cluster at the Physical Research Laboratory, Ahmedabad for performing calculations. V.A.Y acknowledges support by the Russian Federation program for organizing and carrying out scientific investigations. WN thanks G.W.F.\,Drake, Z.C.\,Yan, and R.\,Neugart for stimulating discussions. PS and WN thank K.\,Pachucki for stimulating discussions. 

\end{document}